# Experimental One-Way Quantum Computing

P.Walther[1], K.J.Resch[1], T.Rudolph[2], E.Schenck[1,*], H.Weinfurter[3,4], V.Vedral[1,5,6], M.Aspelmeyer[1] & A.Zeilinger[1,7]

[1] *Institute of Experimental Physics, University of Vienna, Boltzmanngasse 5, 1090 Vienna, Austria*

[2] *QOLS, Blackett Laboratory, Imperial College London, London SW7 2BW, United Kingdom*

[3] *Department of Physics, Ludwig-Maximilians-University, D-80799 Munich, Germany*

[4] *Max-Planck-Institute for Quantum Optics, D-85741 Garching, Germany*

[5] *The Erwin Schrödinger Institute for Mathematical Physics, Boltzmanngasse 9, 1090 Vienna, Austria*

[6] *The School of Physics and Astronomy, University of Leeds, Leeds, LS2 9JT, United Kingdom*

[7] *IQOQI, Institute for Quantum Optics and Quantum Information, Austrian Academy of Sciences, Boltzmanngasse 3, 1090 Vienna, Austria*

[*] *permanent address:* Ecole normale supérieure, 45, rue d´Ulm, 75005 Paris, France

**Standard quantum computation is based on sequences of unitary quantum logic gates which process qubits. The one-way quantum computer proposed by Raussendorf and Briegel is entirely different. It has changed our understanding of the requirements for quantum computation and more generally how we think about quantum physics. This new model requires qubits to be initialized in a**

**highly-entangled cluster state. From this point, the quantum computation proceeds by a sequence of single-qubit measurements with classical feedforward of their outcomes. Because of the essential role of measurement a one-way quantum computer is irreversible. In the one-way quantum computer the order and choices of measurements determine the algorithm computed. We have experimentally realized four-qubit cluster states encoded into the polarization state of four photons. We fully characterize the quantum state by implementing the first experimental four-qubit quantum state tomography. Using this cluster state we demonstrate the feasibility of one-way quantum computing through a universal set of one- and two-qubit operations. Finally, our implementation of Grover's search algorithm demonstrates that one-way quantum computation is ideally suited for such tasks.**

The quantum computer[1,2] is a powerful application of the laws of quantum physics. Such a device will be far more efficient at factoring[3] or database searches[4] compared to its classical counterparts[5]. Considerable effort has been directed toward understanding the role of measurement and entanglement in quantum computation[6-12]. A significant step forward in our understanding was the introduction of the "one-way" quantum computer[13-17] which highlights the importance of both measurement and entanglement in a striking way. In this model, all of the entanglement is provided in advance through a highly-entangled multi-particle cluster state[13]. The quantum computation on the cluster state proceeds via local, single-qubit projective measurements with the outcomes potentially affecting those measurement settings that follow. It is a strength of the cluster state model that the intrinsic randomness of quantum measurement results creates specific types of errors which can be corrected through this classical feedforward. Most importantly, feedforward makes cluster state quantum computation deterministic. In the present, proof-of-principle experiment we perform measurements using fixed single-port polarizers making our computations probabilistic. Different algorithms require only a different pattern of adapted single-qubit operations on a sufficiently large cluster state. Since it is entirely based on single-particle measurements instead of unitary evolution, the computation is inherently not time-reversible - it is one-way. Most importantly cluster state quantum computation is universal[14,17] in that any quantum circuit can be implemented on a suitable cluster state. Open theoretical questions remain about the scalability under realistic noise conditions required for fault-tolerant one-way quantum computation. Although a threshold has been proven to exist[18], it is unknown whether cluster state quantum computation will be more or less sensitive to noise than the standard model.

The one-way quantum computer does not perform quantum logic on the individual qubits of the cluster state. In order to describe the computational circuit, we need to distinguish between the *physical qubits*, in our case the polarization state of photons, which make up the cluster state and on which actual measurements are carried out, and *encoded qubits*, on which the computation is actually taking place. Due to the specific entanglement of the cluster state, no individual physical qubit carries any information about an input state. Therefore, each encoded qubit is written on the cluster state non-locally, i.e. the information is carried by the correlations between the physical qubits. As the quantum computation proceeds, the encoded input qubits are processed in the imprinted circuit, whose output is finally transferred onto physical readout qubits. Interestingly, while the entanglement between the physical qubits in general decreases as a result of the measurement sequence, the entanglement between encoded qubits may increase.

A cluster state can be thought of as emerging from an array of equally prepared independent qubits, which then interact via controlled-phase (CPhase) gates with their nearest (connected) neighbours. Specifically, a cluster state can be built up as follows: A large number of physical qubits are each prepared in the superposition state $|+\rangle = (|0\rangle + |1\rangle)/\sqrt{2}$, where $|0\rangle$ and $|1\rangle$ are the computational basis of the physical qubits. A CPhase operation $|j\rangle|k\rangle \rightarrow (-1)^{jk}|j\rangle|k\rangle$, with ($j, k \in 0, 1$) is then applied between pairs of neighbouring, connected qubits and effectively generates entanglement between them. The choice which physical qubit neighbours are entangled by the CPhase operations, drawn as connecting "bonds" (Figure 1), determines the structure of the cluster state which defines the basic type quantum circuit it can implement. This construction provides an intuitive understanding for the graphical representation of cluster states as connected arrays of physical qubits, in which each line corresponds to a previous nearest-neighbour interaction. We will later demonstrate how the highly entangled cluster states can be generated in a different way, directly from nonlinear optical processes.

Given a cluster state, two basic types of single-particle measurements suffice to operate the one-way quantum computer. Measurements in the computational basis $\{|0\rangle_j, |1\rangle_j\}$ have the effect of disentangling, i.e., removing, the physical qubit $j$ from the cluster leaving a smaller cluster state. Such operations can be used to modify the structure of the cluster and thus the imprinted circuit. The measurements which perform the actual quantum information processing are made in the basis $B_j(\alpha) = \{|+\alpha\rangle_j, |-\alpha\rangle_j\}$ where $|\pm\alpha\rangle_j = (|0\rangle_j \pm e^{i\alpha}|1\rangle_j)/\sqrt{2}$ ($\alpha$ is a real number). The choice of measurement basis determines the single-qubit rotation, $R_z(\alpha) = \exp(-i\alpha\sigma_z/2)$, followed by a Hadamard operation, $H = (\sigma_x + \sigma_z)/\sqrt{2}$, on encoded qubits in the cluster ($\sigma_x, \sigma_y, \sigma_z$ being the usual Pauli matrices). Combinations of rotations about the z-axis and Hadamard operations can implement $R_x(\alpha) = \exp(-i\alpha\sigma_x/2)$ rotations through the matrix identity $R_x(\alpha) = HR_z(\alpha)H$. Any quantum logic operation can be carried out by the correct choice of $B_j(\alpha)$ on a sufficiently large cluster state. We define the outcome $s_j$ of a measurement on the physical qubit $j$ 0 if the measurement outcome is $|+\alpha\rangle_j$ and 1 if the outcome is $|-\alpha\rangle_j$. In those cases where the 0 outcome is found, the computation proceeds as desired.

However, in those cases where the 1 outcome is found, a well-defined Pauli error is introduced. Feedforward, such that the output controls future measurement, compensates for these known errors.

For the implementations of single- and two-qubit quantum logic, we post-select only those cases where the 0 outcome is found and the computation proceeds error free. In the final section, where we report the implementation of Grover's search algorithm, the feedforward determines the final, classical, measurement. There we measured all possible combinations of the measurement results individually and applied the feedforward relation in such a way that the earlier measurements define the physical meaning of the final ones.

Even a small cluster state suffices to demonstrate all the essential features of one-way quantum computing. Each of the three- and four-particle cluster states shown in Figure 1 can implement the quantum circuit shown to its right that consist of a series of single- and two-qubit quantum gates. The computation proceeds via single-particle measurements carried out from the left side of the cluster to the right side, where the final readout takes place. The important feature of the quantum circuits is that the output of one circuit can be fed into the input of a subsequent one if their cluster states are bonded together by CPhase operations. Thus these small circuits, which form a universal set of logic gates, can be used as subunits for a fully-functional quantum computer.

As an example, consider the four-particle box cluster state $\left| \Phi_{\square 4} \right\rangle$ on a 2-dimensional lattice (Figure 1e). The encoded input to the two-qubit quantum circuit is the product state $\left| \Psi_{in} \right\rangle = \left| + \right\rangle_{1E} \left| + \right\rangle_{2E}$, where the numerical subscript labels the qubit and the subscript $E$ is used to distinguish encoded qubits from physical qubits. The circuit processes this pair of encoded qubits through a sequence beginning with a CPhase gate, followed by single-qubit rotations $R_z(-\alpha)$ and $R_z(-\beta)$ on encoded qubits 1 and 2, then a Hadamard operation, $H$, on both qubits, ending with a second CPhase operation. The values for $\alpha$ and $\beta$ of the rotation gates are set by the choice of measurement bases $B_1(\alpha)$ and $B_4(\beta)$ on the physical qubits 1 and 4 respectively. The output of the quantum computation is transferred onto physical qubits 2 and 3. This kind of two-qubit quantum gate is essential for universal quantum computation since it can generate entanglement between the encoded qubits.

On the other hand, by changing the geometry of the cluster state to the one-dimensional cluster $\left|\Phi_{lin4}\right\rangle$ (Figure 1b), it now results in a different quantum circuit corresponding to a set of single-qubit rotations on one encoded qubit. Consecutive measurements $B_1(\alpha)$, $B_2(\beta)$, and $B_3(\gamma)$ on the physical qubits 1, 2, and 3 transform the input state, in our case $\left|\Psi_{in}\right\rangle = \left|+\right\rangle_{1E}$, to $\left|\Psi_{out}\right\rangle = HR_z(-\gamma)R_x(-\beta)R_z(-\alpha)\left|\Psi_{in}\right\rangle$ and store the output on qubit 4.

In order to demonstrate all the circuits shown in Figure 1a-1e, it is sufficient to first produce a linear cluster state of four qubits. The particular circuit implemented is then determined by the order of the measurements performed. Specifically, the one-dimensional linear structure (Figure 1a and 1b) is implemented by sequentially measuring qubits 1, 2 and 3, with the final result then being available at qubit 4. The two-dimensional horseshoe structures (Figure 1c and 1d) are implemented by measuring either qubits 2 and 3 or 1 and 4 with the final result then being available at qubits 1 and 4 or 2 and 3, respectively. The four qubit box cluster (Figure 1e) can be obtained from the four-qubit linear cluster by Hadamard rotations and by swapping (i.e., relabeling) the physical qubits 2 and 3. All this will be described in more detail in the next section and in the section "two-qubit gates".

The difficulty of one-way quantum computer lies with the cluster-state preparation. Cluster states naturally arise in spin chains or spin lattices via nearest-neighbour Ising interaction[13], a well-known interaction model in solid-state physics. Therefore, the first proposals to achieve cluster states were based on analogues in dipole-dipole coupling between atoms in optical lattices[15,19]. Although photon-photon interactions are negligible, recent proposals have nevertheless shown that optical systems may be well-suited for implementing the cluster state model. These schemes utilize sequences of probabilistic quantum-logic gates[8,20-23] based on linear optical elements to construct large photonic cluster states[24,25]. These optical one-way quantum computation proposals are less demanding on resources than the comparable optical implementation in the standard model[8].

In the present work, we have employed nonlinear optics to directly produce four-photon cluster states. This method exploits a mode- and polarization-entangled four-photon state[26] produced in pulsed-pump spontaneous parametric down conversion (SPDC) (see Methods). We reconstructed the density matrix of the four-qubit cluster state using quantum state tomography and studied the state's entanglement properties relevant for quantum computation. We then implement all of the quantum circuits shown in Figure 1a-1e and demonstrate a two-qubit quantum search algorithm. In doing so, we have demonstrated the first universal set of gates and an important algorithm in a one-way quantum computer.

**Creation and Characterization of the Cluster State**

Our cluster state is produced experimentally using the mode- and polarization-entangled output of nonlinear spontaneous parametric down-conversion and linear optical elements as described in detail in the Methods section. When four photons are emitted into the output modes of the polarizing beam-splitters 1, 2, 3, and 4, they are in the highly-entangled cluster state,

$$|\Phi_{cluster}\rangle = \frac{1}{2}\Big(|H\rangle_1 |H\rangle_2 |H\rangle_3 |H\rangle_4 + |H\rangle_1 |H\rangle_2 |V\rangle_3 |V\rangle_4 + |V\rangle_1 |V\rangle_2 |H\rangle_3 |H\rangle_4 - |V\rangle_1 |V\rangle_2 |V\rangle_3 |V\rangle_4\Big), \quad (1)$$

where $|H\rangle$ and $|V\rangle$ represent horizontally- and vertically-polarized photon states and the subscript labels the spatial mode. This state, $|\Phi_{cluster}\rangle$, is equivalent to the four-qubit linear cluster, $|\Phi_{lin4}\rangle$, and the horse-shoe cluster states, $|\Phi_{\sqsubset 4}\rangle$ & $|\Phi_{\sqsupset 4}\rangle$ (Figure 1) under the local unitary operation $H_1 \otimes I_2 \otimes I_3 \otimes H_4$ on the physical qubits, where $H_i$ ($I_i$) is a Hadamard (Identity) operation on qubit $i$. The four-qubit linear cluster can easily be reduced to a three-qubit linear cluster by measuring qubit 1 in the computational basis of the cluster and thus disentangling it from the rest. The state, $|\Phi_{cluster}\rangle$, can be converted to the box cluster state (Figure 1e) by the local unitary operation $H_1 \otimes H_2 \otimes H_3 \otimes H_4$ and a swap (or relabeling) of qubits 2 and 3. Note that the four-qubit cluster state thus realized is also the smallest cluster state that represents a new kind of entanglement[28], while the two-qubit and three-qubit cluster states are locally-equivalent to a maximally-entangled Bell state and the three-qubit GHZ state, respectively.

**State Tomography**

We have completely characterized our state via quantum-state tomography, which is a method for extracting the density matrix of a quantum state, from a discrete set of experimental measurements. While quantum process and state tomography has been performed with up to three qubits[28,29] this is the first time that a four-qubit density matrix has been determined from a complete, experimentally-obtained density matrix. For a four-photon polarization state, like our cluster state, the full density matrix, $\rho$, is a $16 \times 16$ dimensional object that can be reconstructed by linear combinations of 256 linearly-independent four-photon polarization projections. We performed each of these 256 measurements for 600 seconds using all combinations of $\left\{|H\rangle,|V\rangle,|+\rangle,|R\rangle\right\}$, where $|+\rangle=(|H\rangle+|V\rangle)/\sqrt{2}$ and $|R\rangle=(|H\rangle-i|V\rangle)/\sqrt{2}$. A maximum of 127 four-fold coincidence counts in 600 seconds were measured in the case of the setting $|H\rangle_1|H\rangle_2|V\rangle_3|V\rangle_4$. Instead of a direct linear combination of measurement results which can lead to unphysical density matrices due to experimental noise, we employ a maximum-likelihood reconstruction technique[30-32]. The resulting density matrix is shown in Figure 2. The dominant diagonal elements represent the four components $|H\rangle_1|H\rangle_2|H\rangle_3|H\rangle_4$, $|H\rangle_1|H\rangle_2|V\rangle_3|V\rangle_4$, $|V\rangle_1|V\rangle_2|H\rangle_3|H\rangle_4$ and $|V\rangle_1|V\rangle_2|V\rangle_3|V\rangle_4$ as expected from Eq. (1), while off-diagonal elements indicate strong coherences between them. The negative coherences are due to the required $\pi$ phase shift in the amplitude for the $|V\rangle_1|V\rangle_2|V\rangle_3|V\rangle_4$ term. Our reconstructed state is in good qualitative agreement with the target state, $|\Phi_{cluster}\rangle$. This can be quantified by the state fidelity $F=\langle\Phi_{cluster}|\rho|\Phi_{cluster}\rangle=(0.63\pm0.02)$, which is the state vector overlap with the ideal state $|\Phi_{cluster}\rangle$. At present no theoretical results exist concerning the fidelity requirements in cluster state quantum computation, therefore the full implication of the value we found for the fidelity are unclear. Nevertheless, it has been proven that bi-separable four-qubit states cannot have a fidelity greater than 0.5 with our target cluster state. Our measured fidelity is clearly above this threshold and therefore the observed fidelity is proof the fact that the state contains the required four-particle entanglement[34] to a significant degree. Our cluster state is a coherent superposition of four different terms, each from a different physical origin (see methods). The fidelity of our state is not perfect because of partial distinguishability of these terms and because of imperfect phase stability in our setup. We expect that, achieving high fidelity with the target state will become rapidly more difficult as the size of the Hilbert space is increased. However, the acceptable amount of noise per qubit is independent of the size of the cluster, whereas it becomes exponentially small for GHZ states[34]. Thus the problem of noise does not increase for larger clusters[34].

**Entanglement Properties**

One-way cluster state quantum computation is based entirely on the entanglement properties of the physical qubits of the initial cluster state. For three-qubit states, there are only two classes of entanglement typified by the GHZ state[35], $|GHZ\rangle = \left(|H\rangle_1 |H\rangle_2 |H\rangle_3 + |V\rangle_1 |V\rangle_2 |V\rangle_3\right)/\sqrt{2}$, and by the W state[36,37], $|W\rangle = \left(|V\rangle_1 |H\rangle_2 |H\rangle_3 + |H\rangle_1 |V\rangle_2 |H\rangle_3 + |H\rangle_1 |H\rangle_2 |V\rangle_3\right)/\sqrt{3}$. As such, the three-qubit cluster state cannot represent a fundamentally different type of state and is, in fact, a GHZ state. On the other hand, our four-qubit cluster state cannot be converted via local unitary operations to either the four-qubit generalizations of the GHZ or the W state but combines important characteristics of both.

As in a four-qubit GHZ state, one can make local measurements on one or two of the qubits and, with only classical communication, leave the remaining qubits in a three-qubit GHZ or in a two-qubit Bell state both of which can serve as an entanglement resource for quantum communication. For an explicit example, we reconstructed the density matrix of qubits 2, 3, and 4 by considering only the subset of our full tomographic measurements where qubit 1 was successfully measured in the state $|+\rangle_1$. Its fidelity compared to an ideal three-photon GHZ state was $\left(0.60 \pm 0.02\right)$, which is above the local realism threshold[38] of 0.56.

While the loss of one qubit in a GHZ state is already sufficient to completely disentangle the state, $N/2$ particles have to be removed from an $N$-particle cluster state in order to leave a separable state. The cluster states share this *persistency of entanglement*[14] with the W states, which show an even stronger robustness against such decoherence. We demonstrate this property by considering the reduced density matrix of qubits 2, 3, and 4, $\rho_{2,3,4}^{red}$ after ignoring qubit 1, i.e., after tracing out the first qubit. This trace has been implemented by summing the two subsets of our tomographic measurements in which the first qubit was measured in the state $|H\rangle_1$ or $|V\rangle_1$. To test for entanglement in the remaining three-qubit state, we employed the entanglement-witness operator

$$W = \frac{1}{4} I^{\otimes 3} - \frac{1}{2} \left( |H\rangle_2 \langle H|_2 \otimes |\Phi^+\rangle_{3,4} \langle \Phi^+|_{3,4} + |V\rangle_2 \langle V|_2 \otimes |\Phi^-\rangle_{3,4} \langle \Phi^-|_{3,4} \right),$$

where

$Tr(W\rho) \geq 0$ for all separable states, but is negative for some entangled states. Our state gives a value of $\left( -0.115 \pm 0.007 \right)$, which is negative by $16\sigma$ and thus proves that entanglement in qubits 2, 3, and 4 persists even after "loss" of qubit 1. This remaining entanglement is between qubits 3 and 4. The loss of another particle destroys all entanglement and leaves a separable two-qubit state. We test this by calculating the eigenvalues of the partial transpose reduced density matrix of qubits 3 and 4 which are $\lambda_1 = 0.49$ , $\lambda_2 = 0.45$, $\lambda_3 = 0.05$ and $\lambda_4 = 0.008$. These values are all positive and thus fulfil the necessary and sufficient conditions for separability in two-particle systems[39,40].

## A One-Way Quantum Computer

Given a cluster state, quantum gates are implemented on the encoded qubits using only a series of single-qubit measurements and classical feed-forward. In this section, we demonstrate the essentials of cluster-state quantum computation. In the present experiment, we created a cluster state and performed fixed polarization measurements, i.e., projections onto the $|+\alpha\rangle_j$ in the bases, $B_j(\alpha)$, which require no feed-forward or subsequent corrections. This reduces the success rate of the computation by a factor of two for every measurement as compared to ideal, deterministic gate operations but it certainly suffices as a proof-of principle. An important challenge for future work is to implement the fast active switching and logic requirements for one-way quantum computation with feed-forward. We have realized a universal set of quantum logic operations in the form of single-qubit rotations and nontrivial two-qubit gates. In the following sections we characterize the quality of each quantum computation by comparing the measured output state to the ideal using the state fidelity. Interesting avenues for study include full quantum gate characterization using quantum process tomography[41,42] or related measures[43].

**Single-Qubit Rotations**

We start with the one-dimensional four-qubit linear cluster state, $\left|\Phi_{lin4}\right\rangle$, which implements an arbitrary single-qubit rotation gate as shown in Figure 1b. In particular, we perform $B_1(\alpha)$, $B_2(\beta)$ and $B_3(\gamma)$ on the physical qubits in the linear cluster basis. The parameters $\alpha$, $\beta$, $\gamma$ are sufficient to rotate the input qubit to anywhere on the Bloch sphere, or more generally to implement an arbitrary $SU(2)$ single-qubit rotation. This computation rotates the encoded input qubit $\left|\Psi_{in}\right\rangle = \left|+\right\rangle_{1E}$ to the output state $\left|\Psi_{out}\right\rangle = HR_z(-\gamma)R_x(-\beta)R_z(-\alpha)\left|\Psi_{in}\right\rangle$, while the output of this computation is left in the quantum state of qubit 4. We finally characterize physical qubit 4 to verify the performance of the computation using single-qubit quantum state tomography (Table 1a).

The three-qubit linear cluster state $\left|\Phi_{lin3}\right\rangle$ is generated from $\left|\Phi_{lin4}\right\rangle$ by disentangling physical qubit 1 from the cluster. This is achieved by measuring physical qubit 1 in the computational basis for the linear cluster, i.e. $\left\{\left|+\right\rangle_1, \left|-\right\rangle_2\right\}$ in the lab basis. We consider only those cases where we find the "+" outcome. This resulting cluster state implements the quantum circuit in Figure 1a, which is a simpler single-qubit rotation gate with rotations determined by the measurements $B_2(\alpha)$ and $B_3(\beta)$ of the second and the third qubit. This rotates the encoded input qubit $\left|\Psi_{in}\right\rangle = \left|+\right\rangle_{1E}$ to the final state, $\left|\Psi_{out}\right\rangle = R_x(-\beta)R_z(-\alpha)\left|\Psi_{in}\right\rangle$, which is again left on physical qubit 4. The computation is directly implemented by performing single-particle measurements on qubits 1, 2 and 3. The single-qubit output stored on qubit 4 is completely characterized by single-qubit tomography. We compare this single output qubit with both the theoretically expected output and the predicted output from our reconstructed four-particle cluster state density matrix. Figure 3 shows the state of qubit 4 on the Bloch sphere in the lab basis in the ideal (left-hand side) and measured (right-hand side) case for three different measurement settings. These measurement settings were chosen to clearly show the effect of changing a single measurement basis. For the three state vectors shown as 1, 2, and 3, $\alpha$ was set to $\pi/2$, $\pi/4$ and 0, respectively, while $\beta$ was fixed to $-\pi/2$. The state fidelities of these and other measurement settings compared to the ideal gate action are shown in Table 1b.

**Two-Qubit Gates**

In order to perform universal quantum computation nontrivial two-qubit quantum-logic gates[4] are required in addition to single-qubit rotations. Well-known examples of such gates are the controlled-NOT (CNOT) or controlled Phase (CPhase) operations. The crucial trait of these gates is that they can change the entanglement between qubits. Gates of this type can be implemented with a two-dimensional cluster. In Figure 1, we show two-dimensional cluster states, the horseshoe cluster states (Figure 1c and 1d) and the box cluster state (Figure 1e), that satisfy this condition for two-qubit quantum logic. Both of their quantum circuits are comprised of single-qubit rotations and CPhase operations that can generate entanglement between two initially separable encoded qubits. Whether or not entanglement is generated depends on the specific circuit and the initial states. We will give a specific example of a two-qubit quantum computation that does not generate entanglement (in the box cluster) and a second computation that does generate entanglement (in the horseshoe cluster).

The box cluster transforms the two-qubit encoded input state according to

$$\left| \Psi_{out} \right\rangle = CPhase\left( H_1 \otimes H_2 \right)\left[ R_z(-\alpha) \otimes R_z(-\beta) \right] CPhase \left| \Psi_{in} \right\rangle, \qquad (2)$$

where $\alpha$ and $\beta$ are determined by measurements $B_1(\alpha)$ and $B_4(\beta)$ on qubits 1 and 4 respectively. For the isolated box (or horseshoe) cluster state, the encoded qubit input state is the product state $\left| \Psi_{in} \right\rangle = \left| + \right\rangle_{1E} \left| + \right\rangle_{2E}$. Consider the case where photons 1 and 4 are measured to be $|V\rangle_1$ and $|H\rangle_4$ in the lab basis. For the box cluster, this corresponds to measuring the "0" outcome of $B_1(\pi)$ and $B_4(0)$ respectively on physical qubits 1 and 4. According to the box quantum circuit, this should perform the computation $|+\rangle_{1E}|+\rangle_{2E} \to |+\rangle_{1E}|-\rangle_{2E}$ with the resulting product state outcome left on qubits 2 and 3. The experimentally-measured final state of qubits 2 and 3, in the lab basis, are shown as a two-qubit density matrix in Figure 4a. The state has a single dominant diagonal element corresponding to the state $|V\rangle_2|H\rangle_3$ and no off-diagonal elements. Recall the conversion from the box basis to the lab basis requires a swap of qubits 2 and 3 and application of a Hadamard on each qubit. This converts $|+\rangle_{1E}|-\rangle_{2E}$ to $|V\rangle_{1E}|H\rangle_{2E}$ in agreement with our measured density matrix. The fidelity of this measured density matrix of qubits 2 and 3 with the ideal output is $(0.93 \pm 0.01)$. We can quantify the entanglement in the output state using the tangle[44] defined as $\tau = \left[ Max\left(0, \sqrt{\lambda_1} - \sqrt{\lambda_2} - \sqrt{\lambda_3} - \sqrt{\lambda_4}\right)\right]^2$, where $\lambda_i$ are the eigenvalues of the matrix $\rho \Sigma \rho^T \Sigma$ and $\Sigma = \sigma_y \otimes \sigma_y$, and are decreasingly ordered with $\lambda_1$ being the largest. The tangle can range from 0 for separable states to 1 for maximally-entangled states. The tangle of our measured density matrix is $(0.01 \pm 0.01)$ in agreement with the expected value of 0 - no entanglement has been generated in this case. The fidelities of other quantum computations in the box cluster are shown in Table 2a including an example where entanglement is generated.

Two-qubit operations are crucial for quantum computation since they can dependent on the measurement settings generate entanglement. The horseshoe cluster state of Figure 1c performs the following quantum circuit

$$|\Psi_{out}\rangle = \left(H_1 \otimes H_2\right)\left[R_z(-\alpha) \otimes R_z(-\beta)\right]CPhase|\Psi_{in}\rangle \qquad (3)$$

For our input state $|\Psi_{in}\rangle = |+\rangle_{1E}|+\rangle_{2E}$, this circuit always generates maximal entanglement. Consider the case where photons 2 and 3 are both measured in the state $|+\rangle$. This measurement corresponds to the "0" outcome of $B_2(0)$ and $B_3(0)$ on physical qubits 2 and 3 and should perform the transformation $|+\rangle_{1E}|+\rangle_{2E} \rightarrow \frac{1}{\sqrt{2}}\left(|0\rangle_{1E}|+\rangle_{2E} + |1\rangle_{1E}|-\rangle_{2E}\right)$, where the output is maximally entangled.

The experimentally-measured output density matrix of photons 2 and 3 is shown in Figure 4b (right-hand side). For comparison, the ideal output density matrix is also shown in Figure 4b (left-hand side). The two density matrices are qualitatively very similar and, indeed, the state fidelity of our measured state with the ideal state is $(0.84 \pm 0.03)$. The tangle of this output state is $\tau = (0.65 \pm 0.11)$ confirming the generation of entanglement between the logical qubits as a result of the quantum computation. Furthermore, this experimentally measured density matrix implies a maximum CHSH Bell parameter[45] of $S = (2.47 \pm 0.08)$, this is well above the $S = 2$ upper limit for local realistic theories. The fidelities of other quantum computations in the horseshoe cluster are shown in Table 2b.

**Grover's search algorithm**

The excitement over quantum computation is based on just a few algorithms, the most well-known being Shor's[3] factorization algorithm and Grover's search algorithm[6]. The latter is extremely important from a fundamental standpoint since it is provably more efficient than the best classical algorithm and from a practical standpoint since fast searching is central to solving difficult problems. Grover's algorithm has been implemented under the standard model[46,47] both in NMR[48,49] and in optical experiments. Here, we demonstrate a two-qubit implementation of Grover's fast quantum search using the cluster state model.

The goal of a search is to identify one out of $N$ elements of a database. Formally one could consider the database as a black box which labels one of $N$ possible inputs leaving all others unchanged. The challenge, then, is to find that labelled state. The best classical strategy chooses the inputs randomly one by one and checks for the label; it takes, on average, about $N/2$ calculations to find the special input. Quantum parallelism in Grover's algorithm allows for all inputs to be processed simultaneously in superposition while interference enhances the probability of finding the desired outcome in only $O\left(\sqrt{N}\right)$ calculations of the function. In the case of a two-bit function ($N = 4$), the difference is even more dramatic since Grover's algorithm needs only one calculation of the function, whereas classically three evaluations are needed in the worst case, and 2.25 evaluations are needed on average.

The quantum circuit for the two-qubit Grover algorithm is shown in Figure 5a. Two input qubits, 1 and 2, are prepared in the state $\left|+\right\rangle_1\left|+\right\rangle_2$. This is a superposition of all four computational basis states $\left|0\right\rangle\left|0\right\rangle$, $\left|0\right\rangle\left|1\right\rangle$, $\left|1\right\rangle\left|0\right\rangle$ and $\left|1\right\rangle\left|1\right\rangle$. As one of four possibilities, the black box could label the state $\left|0\right\rangle\left|1\right\rangle \rightarrow -\left|0\right\rangle\left|1\right\rangle$ by changing its sign and leaving all other states unchanged. Note that the change of sign is equivalent to a bit flip on an ancillary qubit. Any of the four possibilities can be set by proper choices of the rotation angles $\alpha$ and $\beta$. The output from the black box is processed by an operation which inverts the amplitudes for each computational state about the mean value. This process amplifies the labelled amplitude and reduces the rest. In the two-qubit case, theory predicts that after a single application of this inversion, the computer outputs the labelled state with probability 1.

We can compare this circuit for the two-qubit Grover algorithm to that implemented by the box cluster state in Figure 1e. The Grover algorithm circuit contains extra fixed single-qubit operations, a $\sigma_Z$ followed by a Hadamard transformation, $H$, on each qubit before the readout in the computational basis. Measurement of a final physical qubit in the computational basis after a $\sigma_Z$ followed by $H$ is equivalent to direct measurement of those qubits in the basis $B(\pi)$, i.e., we can absorb these additional fixed single-qubit operations into the readout stage. The quantum circuit implemented by the box cluster can be seen as precisely that one required for Grover's algorithm provided the final readout measurements are made in $B_{2,3}(\pi)$ on physical qubits 2 and 3.

The cluster state computation proceeds as follows. The encoded qubits begin in the state $\left|+\right\rangle_1 \left|+\right\rangle_2$. Setting the measurement angles, $\alpha\beta$, to $\pi\pi$, $\pi0$, $0\pi$, and $00$ determine the black box settings 00, 01, 10, and 11, respectively. In principle these settings remain hidden. The result of a measurement of physical qubit $i$ is $s_i$ which is 0 (1) for a measurement of $\left|+\alpha\right\rangle$ $\left(\left|-\alpha\right\rangle\right)$. For the computation to proceed deterministically, the black box must provide the measurement outcomes for feed-forward. The encoded qubits are transferred nonlocally to the remaining physical qubits in the cluster. Remarkably, the inversion-about–the-mean process is already hard-wired into the structure of the cluster state and is automatically implemented. The output of the computation, including feed-forward (FF), are two bits $\left\{s_2 \oplus s_4, s_3 \oplus s_1\right\}$ identifying the black box.

In Figure 5b, we show the experimental outcomes of the quantum computation. As in the previous computations, measurements were made using quarter-wave plates and polarizers. The "no FF" data are those computational outputs given that the black box outcomes, $s_1$ and $s_4$, were 0 which requires no feed-forward but reduces the success rate to $1/4$. In addition, we have measured individually all possible correlations between the measurement results from the black box and the readout. This enables us to implement the simplest feed-forward possible, where the earlier measurement determines the meaning of the final readout. Thus, when the black box measurement outcomes are other than $s_1 = 0$ and $s_4 = 0$, it is necessary to reinterpret the readout via the bit-wise addition shown above; the "FF" row of data shows the sum of the readouts interpreted in this way. In either case, the probability of the quantum computer determining the correct outcome is about 90%. These high-fidelity results shown in Figure 5b constitute the first demonstration of a quantum algorithm in a cluster state quantum computer.

## Conclusion

We have generated a four-qubit cluster state and characterized that state and its entanglement features. With that cluster, and taking advantage of a curious equivalence of a number of cluster states, we have demonstrated a universal set of single- and nontrivial two-qubit quantum logic gates in a one-way quantum computer. Our final realization of the Grover algorithm strongly underlines the basic simplicity of the cluster state approach. Given the various alternatives for their creation, such as linear optics, ion traps, and optical lattices, and the recent advances in the preparation of multi-particle entangled states, cluster states are promising for inclusion in future implementations of quantum computation. The most important challenges for the optical approach presented here are (a) realization of cluster states on demand, (b) generating cluster states with more qubits and (c) implementation of fast feedforward where the earlier measurement actually changes the setting of a future measurement in real time.

## Methods

### Experimental Preparation of Cluster States

In our experiment, entangled photons are created by using type-II parametric down-conversion[50]. A frequency-doubled laser pulse at 395nm makes two passes through a $\beta$-barium borate (BBO) crystal, which emits highly-entangled photons into the forward pair of modes a & b and backward pair of modes c & d (see Figure 6). To counter the effect of birefringence in the BBO crystal, the polarization in each mode is rotated by 90° and the photons pass through compensation crystals which erase transverse and longitudinal walk-off. Final half wave plates (HWPs), one for each photon pair, and the tilt of the compensation crystals allow for the production of any of the four Bell states. The modes of the forward emitted pairs a & b and the modes of the backward emitted pairs c & d are coherently combined at polarizing beam-splitters (PBSs) by adjusting the position of the delay mirror for the UV-pump. The preparation of the cluster state is based on the simultaneous emission of four photons. The construction of the setup allows for four photon events to come from either two entangled pairs, one forward and one backward, or form double-pair emission into the modes a & b and c & d[26].

Proposed methods for producing cluster states are based on series of two-qubit gates such as the CPhase or CNOT. In our case the four-photon cluster state is generated directly from parametric down-conversion. Because of its intrinsic probabilistic nature the down-conversion process becomes exponentially inefficient for generating larger cluster states. Our way of generating the cluster states furthermore exploits the properties of polarizing beam-splitters (PBSs) and uses post-selection. A PBS is an optical device which transmits horizontally-polarized light and reflects vertically-polarized light. Considering the two-photon case, where after the PBS in each mode one photon has to be detected in each mode after the PBS, both incoming photons must have the same polarization, when they come from different input modes, or they must have orthogonal polarizations when entering along the same input mode. If the source produces simultaneously into the forward pair of modes a $\left|\Phi^-\right\rangle_{a,b}$ state, and backwards a $\left|\Phi^+\right\rangle_{c,d}$ state, only the state $\left|H\right\rangle_1\left|H\right\rangle_2\left|H\right\rangle_3\left|H\right\rangle_4 - \left|V\right\rangle_1\left|V\right\rangle_2\left|V\right\rangle_3\left|V\right\rangle_4$ results in a four-photon coincidence. However, there exists also the case where a four-photon emission into the two modes on either side results in a four-fold coincidence. The state must be in this case $-\left|H\right\rangle_1\left|H\right\rangle_2\left|V\right\rangle_3\left|V\right\rangle_4$ coming from the $\left|\Phi^-\right\rangle_{a,b}$ setting and $+\left|V\right\rangle_1\left|V\right\rangle_2\left|H\right\rangle_3\left|H\right\rangle_4$

coming from the $\left|\Phi^{+}\right\rangle_{c,d}$ setting. The final emerging state is a superposition of all four terms.

In order to end up in a cluster state the phase of the $-\left|H\right\rangle_{1}\left|H\right\rangle_{2}\left|V\right\rangle_{3}\left|V\right\rangle_{4}$ term has to be shifted by $\pi$. This can be done using a HWP in one mode, where according to a rotation by an angle $\phi$ the state after the PBSs evolves to $-\cos(2\phi)\left|H\right\rangle_{1}\left|H\right\rangle_{2}\left|V\right\rangle_{3}\left|V\right\rangle_{4}$. Thus any HWP rotation larger than 22.5° results in a sign flip. At the same time, the Bell state is rotated to $\cos(\phi)\left|\Phi^{-}\right\rangle_{a,b}+\sin(\phi)\left|\Psi^{+}\right\rangle_{a,b}$, where the amount of the wanted $\left|\Phi^{-}\right\rangle_{a,b}$ state is decreased by a factor of $\cos(\phi)$. Due to the intrinsic property of the PBS only the amplitudes for the $\left|H\right\rangle_{1}\left|H\right\rangle_{2}\left|H\right\rangle_{3}\left|H\right\rangle_{4}-\left|V\right\rangle_{1}\left|V\right\rangle_{2}\left|V\right\rangle_{3}\left|V\right\rangle_{4}$ terms are affected, while the $\left|\Psi^{+}\right\rangle_{1,2}$ Bell state does not contribute to any four-fold coincidence.

Note that after each PBS a quarter-wave plate (QWP) was placed to compensate birefringence effects. For each measurement, the phase of the back-reflected pair or four-photon was kept fixed and verified for each measurement setting. Taking into account the emission rates of the source for the entangled pairs, 28000 s[-1] two-photon coincidences for the forward-emitted pair and 18000 s[-1] coincidences for the backward-emitted pair, theoretical calculations show that a HWP rotation by 27.5° results in the maximally entangled cluster state of the form

$$\left|\Phi_{cluster}\right\rangle=\frac{1}{2}\left(\left|H\right\rangle_{1}\left|H\right\rangle_{2}\left|H\right\rangle_{3}\left|H\right\rangle_{4}+\left|H\right\rangle_{1}\left|H\right\rangle_{2}\left|V\right\rangle_{3}\left|V\right\rangle_{4}+\left|V\right\rangle_{1}\left|V\right\rangle_{2}\left|H\right\rangle_{3}\left|H\right\rangle_{4}-\left|V\right\rangle_{1}\left|V\right\rangle_{2}\left|V\right\rangle_{3}\left|V\right\rangle_{4}\right).$$

Thus, a HWP in mode a has been rotated to prepare this state. Fine-tuning has been done by short measurements to obtain approximately equal count-rates for each component.

**Acknowledgements**: The authors thank H. J. Briegel, D. Browne and M. Zukowski for theoretical discussions and C. Först for assistance with graphics. This work was supported by the Austrian Science Foundation (FWF), NSERC, EPSRC, the European Commission under project RAMBOQ, and by the Alexander von Humboldt-Foundation.

**Correspondences** and requests for materials should be addressed to Ph.W. (pwalther@quantum.at) or A.Z. (zeilinger-office@quantum.at)

**Figure 1** Few-qubit cluster states and the quantum circuits they implement. For each three- and four-qubit cluster the quantum state and the operations by its circuit are shown. In our experiment the computational basis $|0\rangle$ and $|1\rangle$ is represented by the horizontal and vertical polarization state, $|H\rangle$ and $|V\rangle$, respectively. The quantum state of those four physical qubits in the box is also explicitly written out as a polarization state where $|H\rangle$ and $|V\rangle$ represent horizontal and vertical polarization states and also represent our computational-basis states $|0\rangle$ and $|1\rangle$ respectively. For the case of the linear clusters **a)** $|\Phi_{lin3}\rangle$ and **b)** $|\Phi_{lin4}\rangle$, consecutive measurements on the qubits (1), 2, and 3 will perform a computation as a series of one-qubit rotation gates. The encoded input state undergoes single-qubit rotations with controllable angles and the output is left on physical qubit 4. In contrast, the horse-shoe clusters **c)** $|\Phi_{\subset4}\rangle$ and **d)** $|\Phi_{\supset4}\rangle$ and the box cluster **e)** $|\Phi_{\square4}\rangle$ form more complex circuits containing both single-qubit *and* two-qubit gates, both of which are necessary to form a universal set of logic gates for quantum computation. In particular, measurements on two of the physical qubits (2 & 3 in the case of $|\Phi_{\subset4}\rangle$, and 1 & 4 in the case of $|\Phi_{\supset4}\rangle$ or $|\Phi_{\square4}\rangle$) will perform the circuit defined by the particular cluster and transfer the logical output onto the remaining two physical qubits (1 & 4 in the case of $|\Phi_{\subset4}\rangle$, and 2 & 3 in the case of $|\Phi_{\supset4}\rangle$ or $|\Phi_{\square4}\rangle$). When these cluster states are not a part of a larger cluster, the encoded input states are always $|\Psi_{in}\rangle = |+\rangle_{1E}$ for the one-qubit gates and $|\Psi_{in}\rangle = |+\rangle_{1E} |+\rangle_{2E}$ for the two-qubit gates. **f)** General input states can be prepared and processed through these operations when these clusters are subunits of larger clusters.

**Figure 2** Density matrix of the four-qubit cluster state in the laboratory basis. Shown are the real (top) and imaginary (bottom) part of the density matrix for the ideal case **a)** and the reconstruction from the experimental four-photon tomography data **b)**. In both cases, there are four large diagonal components

corresponding to HHHH, HHVV, VVHH, and VVVV. The coherences between each of these diagonal elements show up as off-diagonal contributions and are necessary for quantum entanglement. The real density matrix was reconstructed via a maximum likelihood method using four-photon coincidence rates obtained in 256 polarization projections.

**Figure 3** Output Bloch vectors from single qubit rotations using a three-qubit linear cluster $\left|\Phi_{lin3}\right\rangle$. The result of the ideal rotations **a)** are compared with the results of the measured rotations **b)** on the encoded input state $\left|+\right\rangle_{1E}$ for three different choices of measurement bases $B_2(\alpha)$. The output state is written in the laboratory basis. The measurement basis $B_3(\beta)$ for qubit 3 was fixed at $\beta = -\frac{\pi}{2}$. The angle $\alpha$ is set to $\frac{\pi}{2}$, $\frac{\pi}{4}$, and 0 for the Bloch vectors labelled 1, 2, and 3 respectively. These different choices of $\alpha$ result in different rotations about the $\left|L\right\rangle$-axis. The sense of the rotation is shown for decreasing $\alpha$. Our final states, extracted from measured single-qubit tomography, had fidelities of $(0.86 \pm 0.03)$, $(0.85 \pm 0.04)$, and $(0.83 \pm 0.03)$ with respect to the ideal output states. Outcomes for other choices of angles and hence other rotations are shown in Table 1b.

**Figure 4** The output density matrices from two different two-qubit computations. Each density matrix is shown as two bar charts with the upper bar chart depicting the real part of the matrix and the lower chart depicting the imaginary part. In the case **a)**, single-qubit measurements were made on qubit 1 and 4 in the box cluster state $\left|\Phi_{\Box 4}\right\rangle$. The measurement settings were $B_1(\pi)$ and $B_4(0)$ which results in the expected output state $\left|V\right\rangle_2 \left|H\right\rangle_3$ in the lab basis. The measured density matrix has $(0.93 \pm 0.01)$ fidelity with this state and no entanglement. In case **b)** single-qubit measurements in $B_2(0)$ and $B_3(0)$ were made on the horseshoe cluster state $|\Phi_{\subset 4}\rangle$. In this case, the expected output density matrix (left-hand side) and the experimentally-measured density matrix (right-hand side) are shown both in the lab basis. The measured and expected density matrices are in good agreement and this is reflected in the fidelity $(0.84 \pm 0.03)$. We extracted the tangle, a measure of entanglement, from the experimentally-measured density matrix to be $\tau = (0.65 \pm 0.11)$. This conclusively demonstrates that our cluster state quantum computer can generate quantum entanglement necessary for universal quantum computation.

**Figure 5** Grover's algorithm in a cluster state quantum computer. **a)** The quantum circuit implementing Grover's search algorithm for two qubits. The box cluster state implements the quantum circuit shown in Figure 1e. These two circuits perform the equivalent computation when the readout measurements on physical qubits 2 and 3 in the box cluster are carried out in the bases $B_{2,3}(\pi)$. The rotations, and hence black box function, are fixed by the measurement settings $B_1(\alpha)$ and $B_4(\beta)$ on physical qubits 1 and 4. The circuit implements one of the four black-boxes in the search algorithm and processes the output through an inversion-about-the-mean operation. The output, which is the final states of physical qubits 2 and 3, reveals which black box was applied. **b)** The experimentally measured outputs of this quantum computation. The data labelled "no FF" show the computational outputs $\{s_2, s_3\}$ in those cases where the measurements in the black box found the "0" outcome. The data labelled "FF" show the outputs for all individually-measured outcomes from the black

box to which the feed-forward relation $\left\{s_2 \oplus s_4, s_3 \oplus s_1\right\}$ has been applied to the output results. The probability for successful identification of the function is approximately 90% in all cases.

**Figure 6** The experimental setup to produce and measure cluster states. An ultra-violet laser pulse makes two passes through a nonlinear crystal which is aligned to produce entangled photon pairs $\left|\Phi^-\right\rangle_{a,b}$ in the forward direction in modes a & b and $\left|\Phi^+\right\rangle_{c,d}$ in the backward direction in modes c & d. Including the possibility of double-pair emission and the action of the polarizing beam-splitters, the four components of the cluster state can be prepared. The incorrect phase on the HHVV amplitude can easily be changed by using a half-wave plate in mode a. The amplitudes can be equalized by adjusting the relative coupling efficiency of those photon pairs produced in the backward pass as compared to the forward pass. Polarization measurements were carried out in modes 1 through 4 using quarter-wave plates and linear polarizers followed by single-mode fibre-coupled single-photon counting detectors behind 3nm interference filters.

**Table 1** Fidelities of single-qubit rotations for the linear clusters states. The measured fidelities of the rotated output states are compared to the ideal rotation gate for the **a)** four-qubit linear cluster $|\Phi_{lin4}\rangle$ and **b)** three-qubit linear cluster $|\Phi_{lin3}\rangle$ from single-qubit tomography (1QT). For additional comparison we also extract the expected performance from full four-qubit tomography (4QT) based on ideal projective measurements on our reconstructed density matrix. The angles **a)** $\{\alpha,\beta,\gamma\}$ and **b)** $\{\alpha,\beta\}$ determine the set of single-qubit rotations implemented on the encoded qubits and are explained in the text. All output states are given in the lab basis.

**a)**

| $\alpha$ | $\beta$ | $\gamma$ | Output State (lab basis) | Fidelity (1QT) | Fidelity (4QT) |
|---|---|---|---|---|---|
| 0 | 0 | 0 | $|H\rangle$ | 0.97±0.03 | 0.81±0.03 |
| $\pi$ | 0 | $-\dfrac{\pi}{2}$ | $|V\rangle$ | 0.93±0.01 | 0.91±0.02 |
| 0 | $-\dfrac{\pi}{2}$ | $-\dfrac{\pi}{2}$ | $\dfrac{1}{\sqrt{2}}\big(|H\rangle+|V\rangle\big)=|+\rangle$ | 0.58±0.08 | 0.63±0.03 |
| $\dfrac{\pi}{2}$ | 0 | $-\dfrac{\pi}{2}$ | $\dfrac{1}{\sqrt{2}}\big(|H\rangle-|V\rangle\big)=|-\rangle$ | 0.87±0.04 | 0.84±0.03 |
| 0 | $-\dfrac{\pi}{2}$ | -0 | $\dfrac{1}{\sqrt{2}}\big(|H\rangle-i|V\rangle\big)=|R\rangle$ | $0.99^{+0.01}_{-0.06}$ | 0.78±0.03 |
| $\dfrac{\pi}{2}$ | 0 | 0 | $\dfrac{1}{\sqrt{2}}\big(|H\rangle+i|V\rangle\big)=|L\rangle$ | $0.99^{+0.01}_{-0.02}$ | 0.94±0.03 |

**b)**

| $\alpha$ | $\beta$ | Output State (lab basis) | Fidelity (1QT) | Fidelity (4QT) |
|---|---|---|---|---|
| $\frac{\pi}{2}$ | $0$ | $\frac{1}{\sqrt{2}}\left(\lvert H\rangle + i\lvert V\rangle\right) = \lvert L\rangle$ | 0.60±0.05 | 0.73±0.04 |
| $\frac{\pi}{2}$ | $-\frac{\pi}{4}$ | $\frac{1}{\sqrt{2}}\left(\lvert H\rangle - e^{-i\pi/4}\lvert V\rangle\right)$ | 0.74±0.06 | 0.78±0.03 |
| $\frac{\pi}{2}$ | $-\frac{\pi}{2}$ | $\frac{1}{\sqrt{2}}\left(\lvert H\rangle - \lvert V\rangle\right) = \lvert -\rangle$ | 0.86±0.03 | 0.83±0.03 |
| $\frac{\pi}{4}$ | $-\frac{\pi}{2}$ | $\cos\left(\frac{\pi}{8}\right)\lvert H\rangle - \sin\left(\frac{\pi}{8}\right)\lvert V\rangle$ | 0.85±0.04 | 0.87±0.03 |
| $0$ | $-\frac{\pi}{2}$ | $\lvert H\rangle$ | 0.83±0.03 | 0.80±0.02 |
| $\frac{\pi}{4}$ | $-\frac{\pi}{12}$ | $\cos\left(\frac{\pi}{8}\right)\lvert H\rangle - \sin\left(\frac{\pi}{8}\right)e^{i7\pi/12}\lvert V\rangle$ | 0.67±0.05 | 0.75±0.03 |
| $\frac{\pi}{2}$ | $-\frac{\pi}{6}$ | $\frac{1}{\sqrt{2}}\left(\lvert H\rangle - e^{i2\pi/3}\lvert V\rangle\right)$ | 0.65±0.5 | 0.76±0.03 |

**Table 2** Fidelities of the output from two-qubit quantum computations in the **a)** box cluster, $\left|\Phi_{\square 4}\right\rangle$, and **b)** horseshoe cluster, $\left|\Phi_{\subset 4}\right\rangle$. In both cases, the encoded input state $\left|\Psi_{in}\right\rangle = \left|+\right\rangle_{1E}\left|+\right\rangle_{2E}$ evolves to a different output state depending on the settings for $\alpha$ and $\beta$. The fidelities of the two-qubit output state were extracted from two-qubit tomography (2QT) and predicted from the four-qubit tomography (4QT). All output states are written in the laboratory basis.

**a)**

| $\alpha$ | $\beta$ | *Output State (lab basis)* | *Fidelity (2QT)* | *Fidelity (4QT)* |
|---|---|---|---|---|
| 0 | 0 | $\left|H\right\rangle_2\left|H\right\rangle_3$ | 0.86±0.02 | 0.80±0.02 |
| $\pi$ | 0 | $\left|V\right\rangle_2\left|H\right\rangle_3$ | 0.93±0.01 | 0.93±0.01 |
| $\pi$ | $\pi$ | $\left|V\right\rangle_2\left|V\right\rangle_3$ | 0.93±0.01 | 0.90±0.01 |
| $\pi$ | $\dfrac{\pi}{2}$ | $\left|V\right\rangle_2\left|L\right\rangle_3$ | 0.94±0.01 | 0.84±0.02 |
| $\dfrac{\pi}{2}$ | $\dfrac{\pi}{2}$ | $\dfrac{1}{\sqrt{2}}\left(\left|V\right\rangle_2\left|R\right\rangle_3 - i\left|H\right\rangle_2\left|L\right\rangle_3\right)$ | 0.64±0.05 | 0.64±0.02 |

**b)**

| $\alpha$ | $\beta$ | *Output State (lab basis)* | *Fidelity (2QT)* | *Fidelity (4QT)* |
|---|---|---|---|---|
| 0 | 0 | $\dfrac{1}{\sqrt{2}}\left(\left|H\right\rangle_1\left|+\right\rangle_4 + \left|V\right\rangle_1\left|-\right\rangle_4\right)$ | 0.84±0.03 | 0.77±0.02 |
| 0 | $-\dfrac{\pi}{2}$ | $\dfrac{1}{\sqrt{2}}\left(\left|H\right\rangle_1\left|L\right\rangle_4 + \left|V\right\rangle_1\left|R\right\rangle_4\right)$ | 0.54±0.03 | 0.64±0.02 |
| $-\dfrac{\pi}{2}$ | 0 | $\dfrac{1}{\sqrt{2}}\left(\left|H\right\rangle_1\left|+\right\rangle_4 + i\left|V\right\rangle_1\left|-\right\rangle_4\right)$ | 0.76±0.03 | 0.73±0.02 |
| $-\dfrac{\pi}{2}$ | $-\dfrac{\pi}{2}$ | $\dfrac{1}{\sqrt{2}}\left(\left|H\right\rangle_1\left|L\right\rangle_4 + i\left|V\right\rangle_1\left|R\right\rangle_4\right)$ | 0.66±0.04 | 0.62±0.02 |

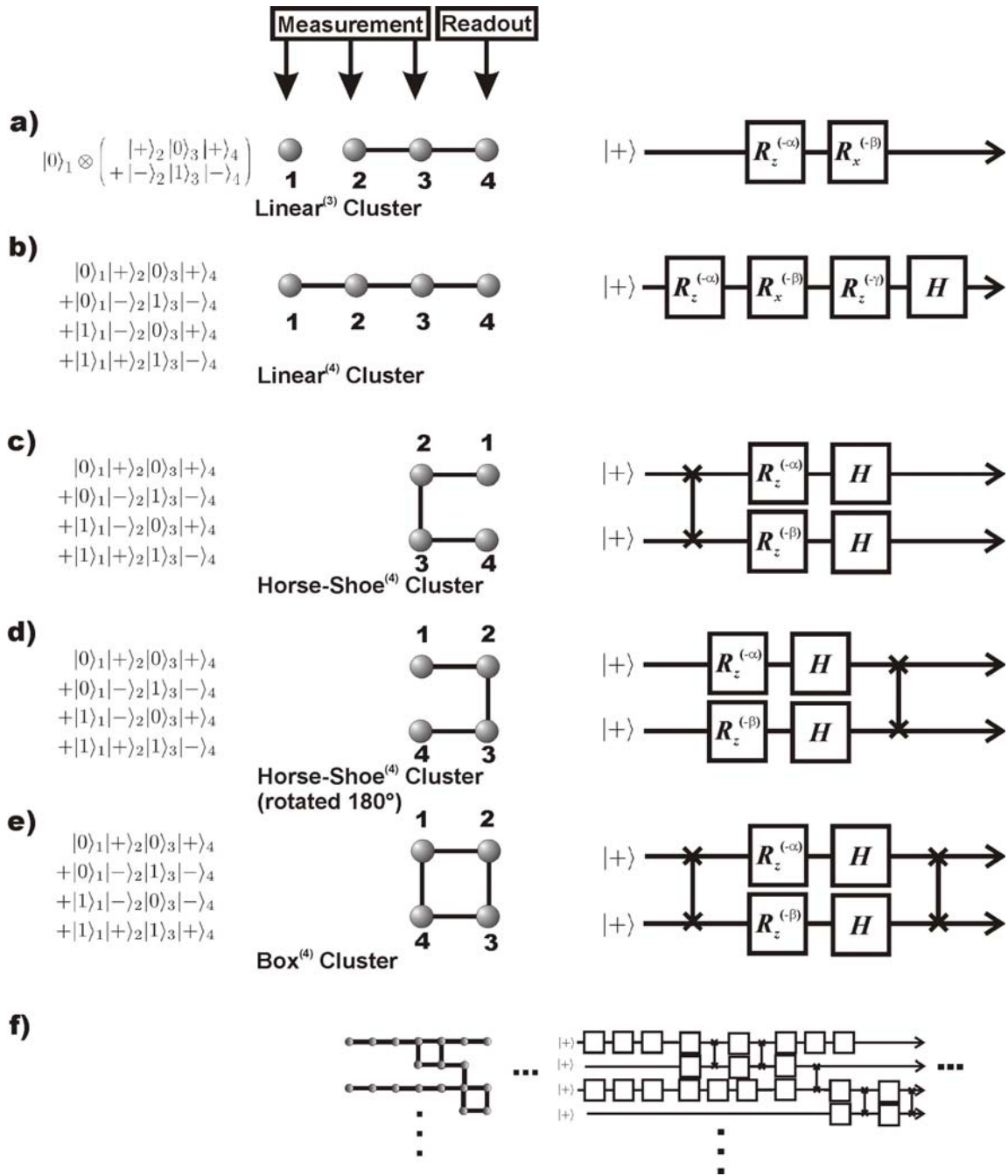

**Figure 1**

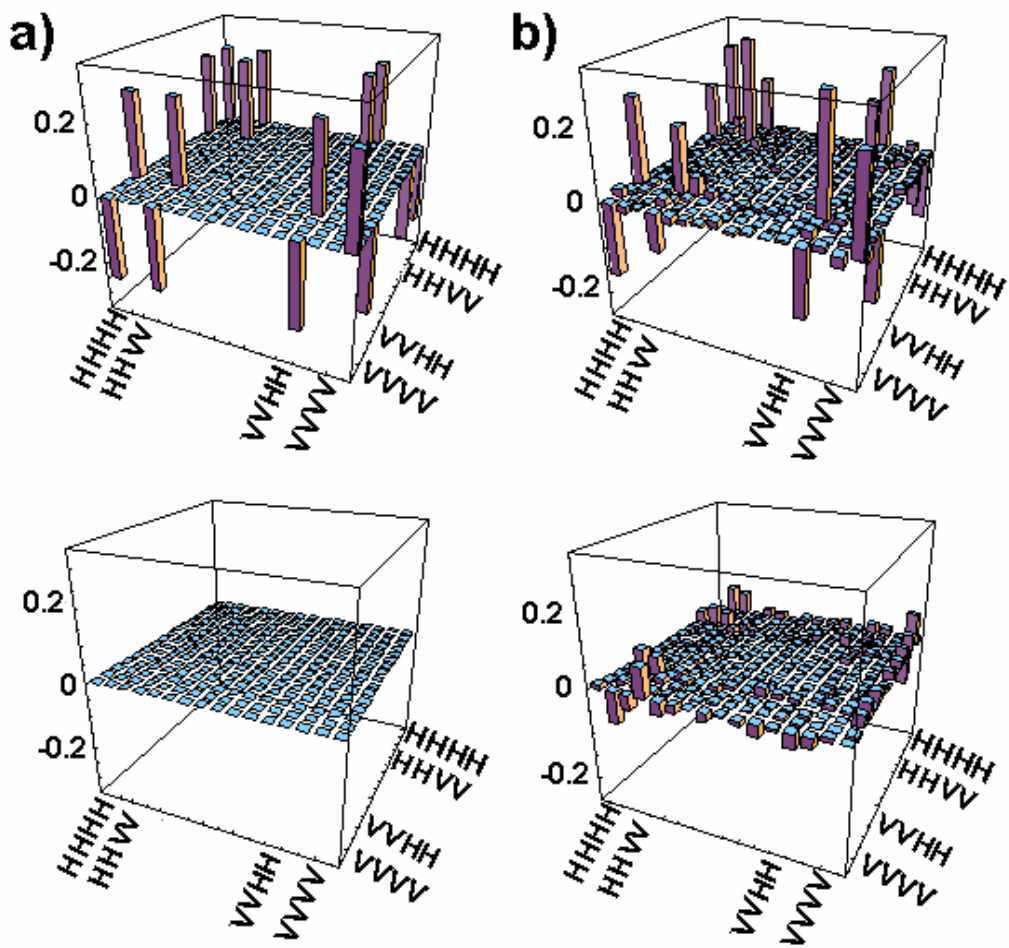

**Figure 2**

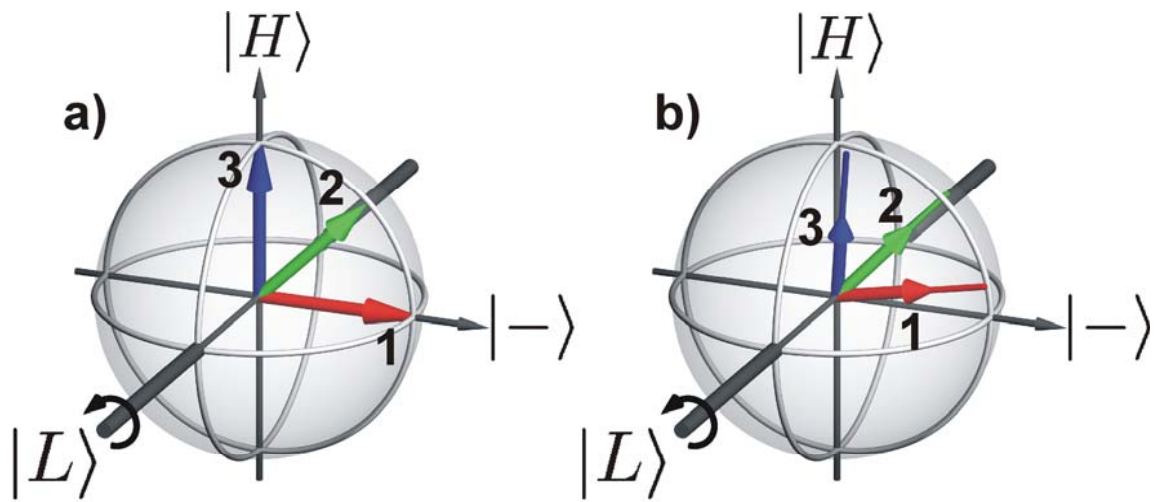

**Figure 3**

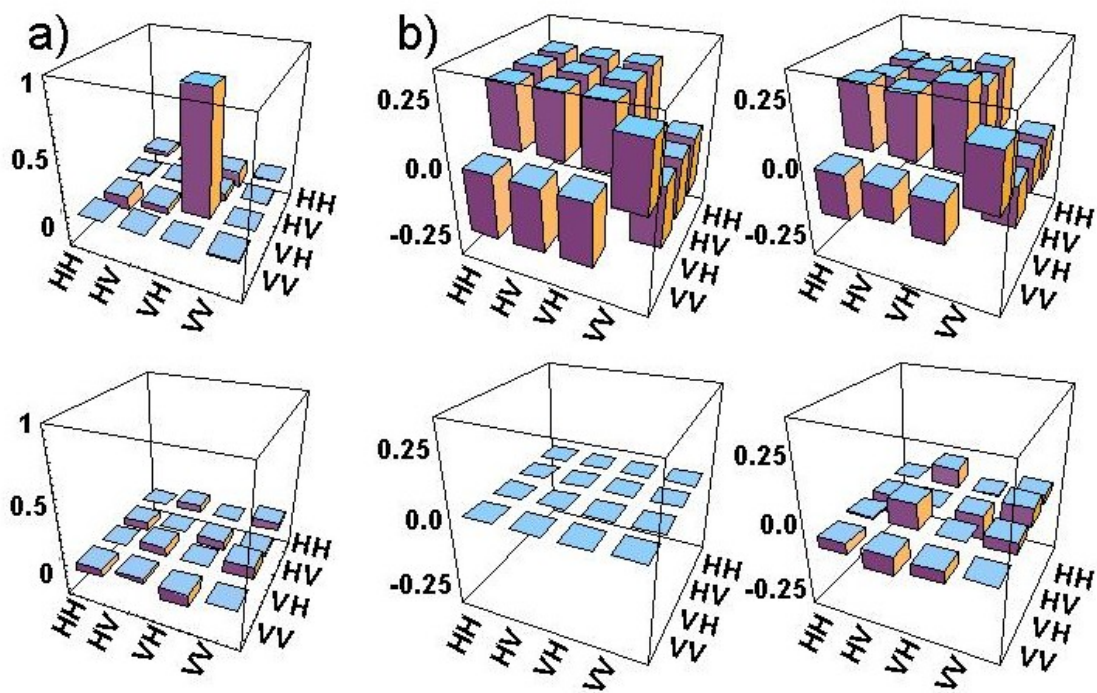

**Figure 4**

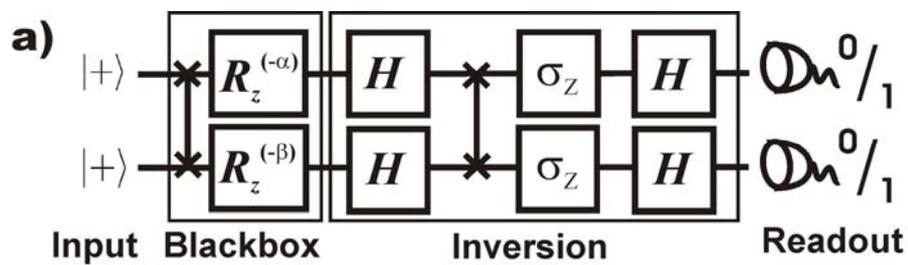

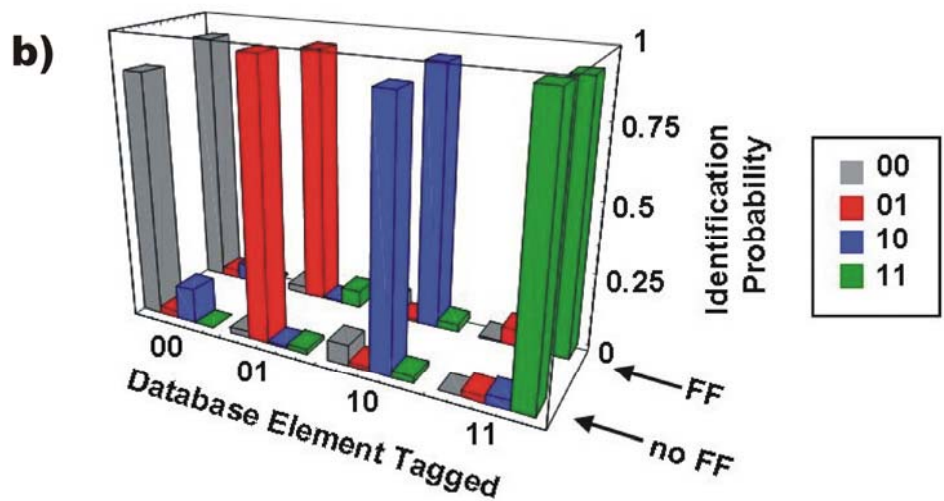

**Figure 5**

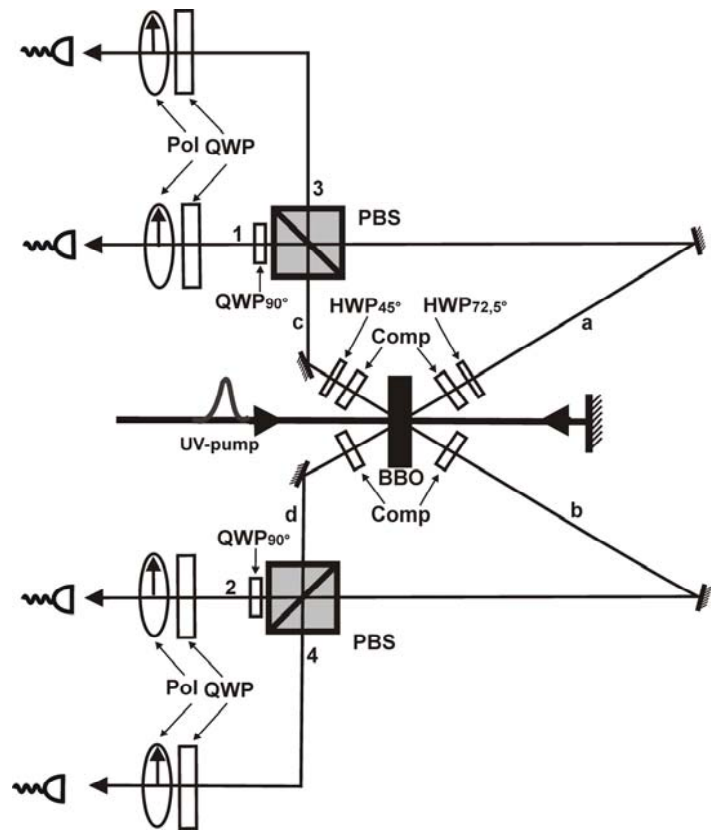

**Figure 6**